\documentclass[prd,twocolumn,showpacs,amsmath,amssymb]{revtex4}
\usepackage{overpic,graphicx}
\usepackage{dcolumn}
\usepackage{bm}
\usepackage{rotating}

\begin{document}
\normalsize
\parskip=5pt plus 1pt minus 1pt

\title{ \boldmath Search for Baryonic Decays of $\psi(3770)$ and $\psi(4040)$}
\author{\small{
M.~Ablikim$^{1}$, M.~N.~Achasov$^{6,a}$, O.~Albayrak$^{3}$, D.~J.~Ambrose$^{39}$, F.~F.~An$^{1}$, Q.~An$^{40}$, J.~Z.~Bai$^{1}$, R.~Baldini Ferroli$^{17A}$, Y.~Ban$^{26}$, J.~Becker$^{2}$
% ...
\\
\vspace{0.2cm}
(BESIII Collaboration)\\
\vspace{0.2cm} {\it
$^{1}$ Institute of High Energy Physics, Beijing 100049, People's Republic of China\\
$^{2}$ Bochum Ruhr-University, D-44780 Bochum, Germany\\
$^{3}$ Carnegie Mellon University, Pittsburgh, Pennsylvania 15213, USA\\
% ...
$^{6}$ G.I. Budker Institute of Nuclear Physics SB RAS (BINP), Novosibirsk 630090, Russia\\
% ...
$^{17}$ (A)INFN Laboratori Nazionali di Frascati, I-00044, Frascati, Italy; (B)INFN and University of Perugia, I-06100, Perugia, Italy\\
% ...
$^{26}$ Peking University, Beijing 100871, People's Republic of China\\
% ...
$^{39}$ University of Rochester, Rochester, New York 14627, USA\\
$^{40}$ University of Science and Technology of China, Hefei 230026, People's Republic of China\\
% ...
$^{a}$ Also at the Novosibirsk State University, Novosibirsk, 630090, Russia\\
% ...
}}}

\vspace{0.4cm}

\begin{abstract}

By analyzing data samples of $2.9$ fb$^{-1}$ collected at $\sqrt
s=3.773$ GeV, $482$ pb$^{-1}$ collected at $\sqrt s=4.009$ GeV and
$67$ pb$^{-1}$ collected at $\sqrt s=3.542$, 3.554, 3.561, 3.600 and
3.650 GeV with the BESIII detector at the BEPCII storage ring, we
search for $\psi(3770)$ and $\psi(4040)$ decay to baryonic final
states, including $\Lambda \bar\Lambda\pi^+\pi^-$, $\Lambda
\bar\Lambda\pi^0$, $\Lambda \bar\Lambda\eta$, $\Sigma^+ \bar\Sigma^-$,
$\Sigma^0 \bar\Sigma^0$, $\Xi^-\bar\Xi^+$ and $\Xi^0\bar\Xi^0$
decays. None are observed, and upper limits are set at the 90\% confidence
level.

\end{abstract}

\pacs{13.25.Gv, 12.38.Qk, 14.40.Gx}

\maketitle

\section{\boldmath Introduction}

Above $D\bar D$ threshold, there are several broad $c\bar c$ resonance
peaks, namely $\psi(3770)$, $\psi(4040)$, $\psi(4160)$ and
$\psi(4415)$. It is important to study the properties of these excited
$J^{PC}=1^{--}$ charmonium states.  The $\psi(3770)$ and $\psi(4040)$
resonances decay quite abundantly into open-charm final states. While
charmless decays of the $\psi(3770)$ and $\psi(4040)$ are possible,
their branching fractions are supposed to be highly suppressed.

Unexpectedly, the BES Collaboration measured the branching fraction
for $\psi(3770)$ decay to non-$D\bar D$ to be $(15\pm5)\%$ by
utilizing varied methods
\cite{plb659_74,prd76_122002,prl97_121801,plb641_145} under the
hypothesis that only one simple $\psi(3770)$ resonance exists in the
center-of-mass energy region from 3.70 to 3.87 GeV.  Meanwhile, the
CLEO Collaboration obtained the branching fraction $\mathcal
B(\psi(3770)\to$ non-$D\bar D)=(-3.3\pm1.4^{+6.6}_{-4.8})$\%, which
corresponds to $\mathcal B(\psi(3770)\to$ non-$D\bar D)<$ 9\% at the
90\% Confidence Level (C.L.) when considering only physical (positive)
values \cite{prl104_159901}. The results are obtained under the
assumption that the interference of the resonance decay,
$\psi(3686)\to\gamma^*\to q\bar q\to hadrons$ with the continuum
annihilation, $\gamma^*\to q\bar q\to hadrons$, is destructive at
$\sqrt s=3.671$ GeV and constructive at $\sqrt s=3.773$ GeV
\cite{prl96_092002}. Since a large non-$D\bar D$ component conflicts
with the theoretical prediction \cite{prl34_369,prd21_203}, it is
important to identify the non-$D\bar D$ decays of the $\psi(3770)$,
which will place the large non-$D\bar D$ component on a solid footing
and shed light on the nature of the $\psi(3770)$.

The BES Collaboration observed the first non-$D\bar D$ decay,
$\psi(3770)\to\pi^{+}\pi^{-}J/\psi$, with a branching fraction of
$(0.34\pm0.14\pm0.09)\%$ \cite{plb605_63}. The CLEO Collaboration
confirmed the same hadronic transition~\cite{prl96_082004}, and
observed other hadronic transitions $\pi^{0}\pi^{0}J/\psi$, $\eta
J/\psi$ \cite{prl96_082004}, and radiative transitions
$\gamma\chi_{cJ}(J=0,1)$ \cite{prl96_182002,prd74_031106} to
lower-lying charmonium states, and the decay to light hadrons
$\phi\eta$ \cite{prd73_012002}. While BES and CLEO have
continued to search for exclusive non-$D\bar D$ decays of
$\psi(3770)$, the total non-$D\bar D$ exclusive components are
less than 2\% \cite{pdg2012}, which motivates the search for other
exclusive non-$D\bar D$ final states.

The $\psi(4040)$ is generally considered to be the 3$^3S_1$ charmonium
state. Studies of its charmless decays are also interesting, and there
are fewer experimental measurements of the branching fractions for
$\psi(4040)$ decay.  The BESIII Collaboration observed the first
production of $e^+e^-\to\eta J/\psi$ at $\sqrt s=4.009$ GeV.  Assuming
the $\eta J/\psi$ signal is from a hadronic transition of the
$\psi(4040)$, the fractional transition rate is determined to be
$\mathcal B(\psi(4040)\to\eta J/\psi)=
(5.2\pm0.5\pm0.2\pm0.5)\times10^{-3}$ \cite{prd86_071101}.  Searching
for other exclusive non-$D\bar D$ decays of $\psi(4040)$ is also
urgently needed.

Since $D$ mesons are not sufficiently massive to decay to baryon
pairs, modes with baryons would be unambiguous evidence for non-$D\bar
D$ decays of $\psi(3770)$. Further, no searches for baryonic decays of
$\psi(4040)$ exist.  In this article, we report results of searches
for baryonic decays of $\psi(3770)$ and $\psi(4040)$, including final
states with baryon pairs ($\Sigma^+ \bar\Sigma^-$, $\Sigma^0
\bar\Sigma^0$, $\Xi^-\bar\Xi^+$, $\Xi^0\bar\Xi^0$) and other $B\bar
B$X modes ($\Lambda \bar\Lambda\pi^+\pi^-$, $\Lambda
\bar\Lambda\pi^0$, $\Lambda \bar\Lambda\eta$).

\section{\boldmath Experiment and data samples}
The data samples used in this analysis were collected at the
$\psi(3770)$ resonance ($\sqrt s=3.773$ GeV), the $\psi(4040)$
resonance ($\sqrt s=4.009$ GeV) and the surrounding continuum ($\sqrt
s=3.542$, 3.554, 3.561, 3.600 and 3.650 GeV), in $e^{+}e^{-}$
collisions produced by the Beijing Electron Positron Collider II
(BEPCII) and acquired with the BESIII detector. BESIII/BEPCII
\cite{bes3} is the major upgrade of BESII/BEPC \cite{bes2} for study
of hadron spectroscopy and $\tau$-charm physics
\cite{bes3_physics}. BEPCII is a double-ring $e^+e^-$ collider
designed for a peak luminosity of $10^{33}$ cm$^{-2}$s$^{-1}$ at a
beam current of $0.93$ A at the $\psi(3770)$ peak. The BESIII detector
with a solid angle coverage of 93\% of 4$\pi$ consists of the
following components: (1) A small cell, helium-based main drift
chamber (MDC) with 43 layers, providing an average single wire
resolution of 135 $\rm \mu$m, a dE/dx resolution that is better than
6\%, and a momentum resolution of 0.5\% for 1 GeV/$c$ charged
particles in the 1.0 Tesla magnetic field; (2) An Electro-Magnetic
Calorimeter (EMC) consisting of 6240 CsI(Tl) crystals arranged in a
cylindrical structure (barrel) and two end caps. The energy resolution
for photons with an energy of 1.0 GeV is 2.5\% (5.0\%) in the barrel
(end caps), and the position resolution is $6$ mm ($9$ mm) in the
barrel (end caps); (3) A Time-of-Flight (TOF) system for particle
identification (PID) composed of two layers (one layer) of
scintillator with time resolution of $80$ ps ($110$ ps) in the barrel
(end caps), corresponding to a $K/\pi$ separation by more than
2$\sigma$ for momenta below about 1 GeV/$c$; (4) These components are
all enclosed in a superconducting solenoidal magnet providing a 1.0
Tesla magnetic field. (5) A muon chamber system (MUC) consisting of
$1000$ m$^2$ of resistive plate chambers (RPC) arranged in 9 layers in
the barrel and 8 layers in the end caps with spatial resolution of $2$
cm.

The integrated luminosity ($\mathcal L$) of the data sets is measured
by using large angle bhabha scatter events.  The data sets for this
analysis consist of $\mathcal L = 2.9$ fb$^{-1}$ of $e^+e^-$
annihilation data collected at the center-of-mass energy of 3.773 GeV,
the peak of the $\psi(3770)$ resonance, 482 pb$^{-1}$ data taken at
the center-of-mass energy of 4.009 GeV, near the peak of the
$\psi(4040)$ resonance and continuum data, which is used to determine
the non-resonant continuum background subtraction, consisting of 23
pb$^{-1}$ taken at center-of-mass energies of 3.542, 3.554, 3.561,
3.600 GeV and 44 pb$^{-1}$ taken at the center-of-mass energy of 3.650
GeV.

The evaluation of detection efficiency, the optimization of the event
selection and the estimation of physics backgrounds are achieved with
simulated Monte Carlo (MC) samples. A GEANT4-based detector simulation
software BOOST \cite{BOOST} includes the geometric and material
description of the BESIII detectors, the detector response, the
digitization models, as well as the tracking of the detector running
conditions and performances. Signal MC samples of $\psi(3770)$ and
$\psi(4040)$ decay to baryonic final states containing 50 000 events
for each channel at $\sqrt s=3.773$ and 4.009 GeV are simulated by
using the generator of KKMC \cite{KKMC}, which includes initial state
radiation (ISR).  For the study of $\psi(3770)$ decay backgrounds, MC
samples of $e^+e^-\to\gamma^{ISR}J/\psi$, $\gamma^{ISR}\psi(3686)$
equivalent to 1.5 times that of the data, and
$e^+e^-\to\psi(3770)\to D\bar D$ and non-$D\bar D$ already
measured experimentally \cite{pdg2012} equivalent to 5.0 times
that of the data are generated.  For the study of $\psi(4040)$
decay backgrounds, about 1 fb$^{-1}$ inclusive ISR MC samples
(mainly $e^+e^-\to\gamma^{ISR}J/\psi$, $\gamma^{ISR}\psi(3686)$ and
$\gamma^{ISR}\psi(3770)$) and $\psi(4040)$ direct decays (mainly open
charm, hadronic and radiative transition production) equivalent
to 2.1 times that of the data are generated.  A scale factor $f_{co}$,
which is used to normalize the continuum products to
$\psi(3770)$/$\psi(4040)$ data, is determined by the integrated
luminosities of the data sets corrected for an assumed $1/s$
dependence of the cross section. We also account for the small
difference in efficiency between the $\psi(3770)$/$\psi(4040)$
data and continuum data. Therefore, MC samples of $e^+e^-\to$ baryonic
final states containing 50 000 events for each mode at $\sqrt
s=3.773$, 4.009, 3.542, 3.554, 3.561, 3.600 and 3.650 GeV are also
generated.  The known decay modes of the charmonium
states are produced by EVTGEN \cite{EvtGen} with branching fractions
being set to world average values \cite{pdg2012} and the unknown ones
by LUNDCHARM \cite{LundCharm}.

\section{\boldmath Event Selection}
The analysis approach and selection criteria are as follows.
Normal requirements are used to select charged particles
reconstructed in the tracking system and photon candidates
reconstructed in the electro-magnetic calorimeter (EMC). 
Charged tracks in BESIII are reconstructed from the main drift
chamber (MDC) hits with good helix fits, which satisfy
$|\rm{cos\theta}|<0.93$, where $\theta$ is the polar angle
with respect to $e^+$ direction. The charged tracks
used in reconstructing $\Lambda$, $\Sigma^+$, $\Sigma^0$,
$\Xi^-$ and $\Xi^0$ decays are not required to satisfy a primary
vertex requirement. Particle identification is used for each
charged particle candidate. We use the combined energy
loss in the drift chamber (dE/dx) and time-of-flight (TOF)
information to compute the particle identification (PID)
confidence levels ($CL_{\pi,K,p}$) for the hypotheses
that the charged track is a $\pi$, $K$ or $p$. We assign
the track to be the $\pi$ with the requirement of
$CL_{\pi}>CL_{K}$, or to be the $p$ with the requirement of
$CL_p>0.001$, $CL_p>CL_{\pi}$ and $CL_p>CL_K$.
We require tracks of proton and anti-proton to have
transverse momenta $p_{xy}>300$ MeV/$c$ due to
differences in the detection efficiencies between data
and Monte Carlo simulation for low-momentum protons and
anti-protons.

Electromagnetic showers are reconstructed
from clusters of energy deposits in the EMC. Efficiency
and energy resolution are improved by adding the energy
deposits in nearby TOF counters. Good photon candidates
are required to satisfy that a shower with an energy
deposited in the barrel region ($|\rm{cos\theta}|<0.8$)
is at least 25 MeV, or at least 50 MeV in the end caps
region ($0.86<|\rm{cos\theta}|<0.92$). To suppress
showers generated by charged particles, the angle
between the photon and the closest charged track is
required to be greater than $10^\circ$. Requirements
on the EMC cluster hit timing are used to suppress
electronic noise and energy deposits unrelated to
the event.

We identify intermediate states through the following
decays: $\Lambda\to p\pi^{-}$, $\pi^{0}\to\gamma\gamma$,
$\eta\to\gamma\gamma$, $\Sigma^{+}\to p\pi^{0}$
($\pi^{0}\to\gamma\gamma$), $\Sigma^{0}\to\Lambda\gamma$
($\Lambda\to p\pi^{-}$), $\Xi^{-}\to \Lambda\pi^{-}$
($\Lambda\to p\pi^{-}$), $\Xi^{0}\to \Lambda\pi^{0}$
($\pi^{0}\to\gamma\gamma$, $\Lambda\to p\pi^{-}$).
For $\Lambda\to p\pi^{-}$, a vertex fit of $p$ and
$\pi^-$ trajectories to a common vertex separated from
the $e^+e^-$ interaction point is made. To eliminate
random $p\pi^-$ combinations, the secondary vertex
fit algorithm is applied to impose the kinematic
constraint between the production and decay vertex with
the run-by-run averaged interaction point
and the fitted $p$ and $\pi^-$ vertex information. For baryon
pair modes ($\Sigma^0 \bar\Sigma^0$, $\Xi^-\bar\Xi^+$,
$\Xi^0\bar\Xi^0$), we only employ a vertex fit of $p$ and
$\pi^-$ to reconstruct $\Lambda$. A loose requirement for
the invariant mass of $p\pi^-$ to be in the range
$|M(p\pi^-)-M(\Lambda)|<40$ MeV/$c^{2}$ is used for all
the modes containing $\Lambda$ in order to improve the
efficiency, where $M(\Lambda)$ is the known mass of
$\Lambda$ \cite{pdg2012}. For $\Xi^{-}\to\Lambda\pi^{-}$,
a vertex fit of $\Lambda$ and $\pi^-$ to a common
vertex is also made.

\begin{figure}[htbp]
\includegraphics[width=8.6cm,height=9.0cm] {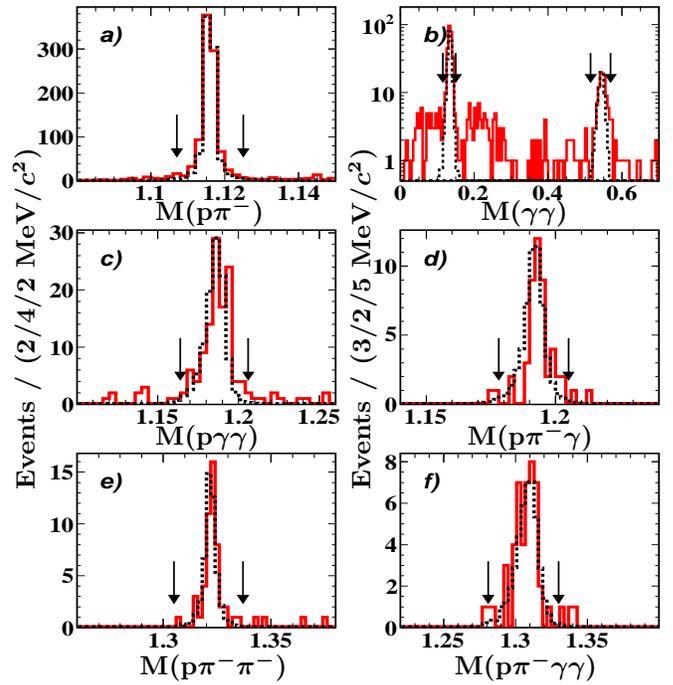}
 \put(-193,172){\normalsize \bf \boldmath M(p$\pi^-$) }
 \put(-193,87 ){\normalsize \bf \boldmath M(p$\gamma\gamma$) }
 \put(-199,0  ){\normalsize \bf \boldmath M(p$\pi^-\pi^-$) }
 \put(-70,172){\normalsize \bf \boldmath M($\gamma\gamma$) }
 \put(-75,87 ){\normalsize \bf \boldmath M(p$\pi^-\gamma$) }
 \put(-80,0  ){\normalsize \bf \boldmath M(p$\pi^-\gamma\gamma$) }
 \put(-247,70){\rotatebox{90}{\normalsize \bf \boldmath Events / (2/4/2 MeV/$c^2$)}}
 \put(-120,70){\rotatebox{90}{\normalsize \bf \boldmath Events / (3/2/5 MeV/$c^2$)}}
 \caption{ \label{Inv_Mass}
  Invariant mass distributions for intermediate states in units of GeV/$c^2$.
  Pairs of arrows indicate the signal region. Solid histogram:
  data at $\sqrt s=3.773$ GeV, dashed histogram: signal
  MC, arbitrary normalization. (a) $\Lambda\to p\pi^-$,
  (b) $\pi^0\to\gamma\gamma$ and $\eta\to\gamma\gamma$ with log scale,
  (c) $\Sigma^{+}\to p\pi^{0}$ ($\pi^{0}\to\gamma\gamma$),
  (d) $\Sigma^{0}\to\Lambda\gamma$ ($\Lambda\to p\pi^{-}$),
  (e) $\Xi^{-}\to \Lambda\pi^{-}$ ($\Lambda\to p\pi^{-}$), and
  (f) $\Xi^{0}\to \Lambda\pi^{0}$ ($\pi^{0}\to\gamma\gamma$,
      $\Lambda\to p\pi^{-}$). }
\end{figure}

For each mode, the reconstructed events passing the above
selection criteria are subjected to a four constraint (4-C)
kinematic fit to make use of momentum and energy
conservation between the initial state ($e^+e^-$ beams)
and the final states. The charged or neutral tracks
comprising these events each have several combinations
to pass through the four constraint kinematic fit,
and only the combination with the smallest $\chi^{2}_{4-C}$,
the $\chi^2$ of the 4-C kinematic fit, is retained for
further study. We require $\chi^{2}_{4-C}<60$ in order to
suppress the backgrounds and improve the signal-to-background
ratio. 

For the final states with two $\pi^0$s, because there exist several
combinations for four $\gamma$s to form the two $\pi^0$s, the candidate
events with the minimum R($\pi^0$), where $R(\pi^0)=
\sqrt{(M(\gamma\gamma)_1-M(\pi^0))^2+(M(\gamma\gamma)_2
-M(\pi^0))^2}$, are selected for further analysis. Here, $M(\pi^0)$ is
the known mass of $\pi^0$ \cite{pdg2012}.  For the baryon pair modes
$\Sigma^+\bar\Sigma^-$, $\Sigma^0\bar\Sigma^0$ and $\Xi^0\bar\Xi^0$,
since they are formed by states of $p\bar p\pi^0\pi^0$,
$\Lambda\bar\Lambda\gamma\gamma$ and $\Lambda\bar\Lambda\pi^0\pi^0$,
there are also multiple solutions to make up the baryon pairs, and we
select the minimum value of $R(j)=\sqrt{(M(i)-M(j))^2+(M(i')-M(j))^2}$
as the optimized one, where $i$ denotes $p\pi^0$, $\Lambda\gamma$ and
$\Lambda\pi^0$, $i'$ is $\bar p\pi^0$, $\bar\Lambda\gamma$ and
$\bar\Lambda\pi^0$, $j$ means $\Sigma^+$, $\Sigma^0$ and $\Xi^0$,
$M(i)$ and $M(i')$ are the invariant mass of $i$ and $i'$, $M(j)$ is
the known mass of $j$ \cite{pdg2012}.

\begin{figure}[htbp]
\includegraphics[width=8.6cm,height=8.6cm] {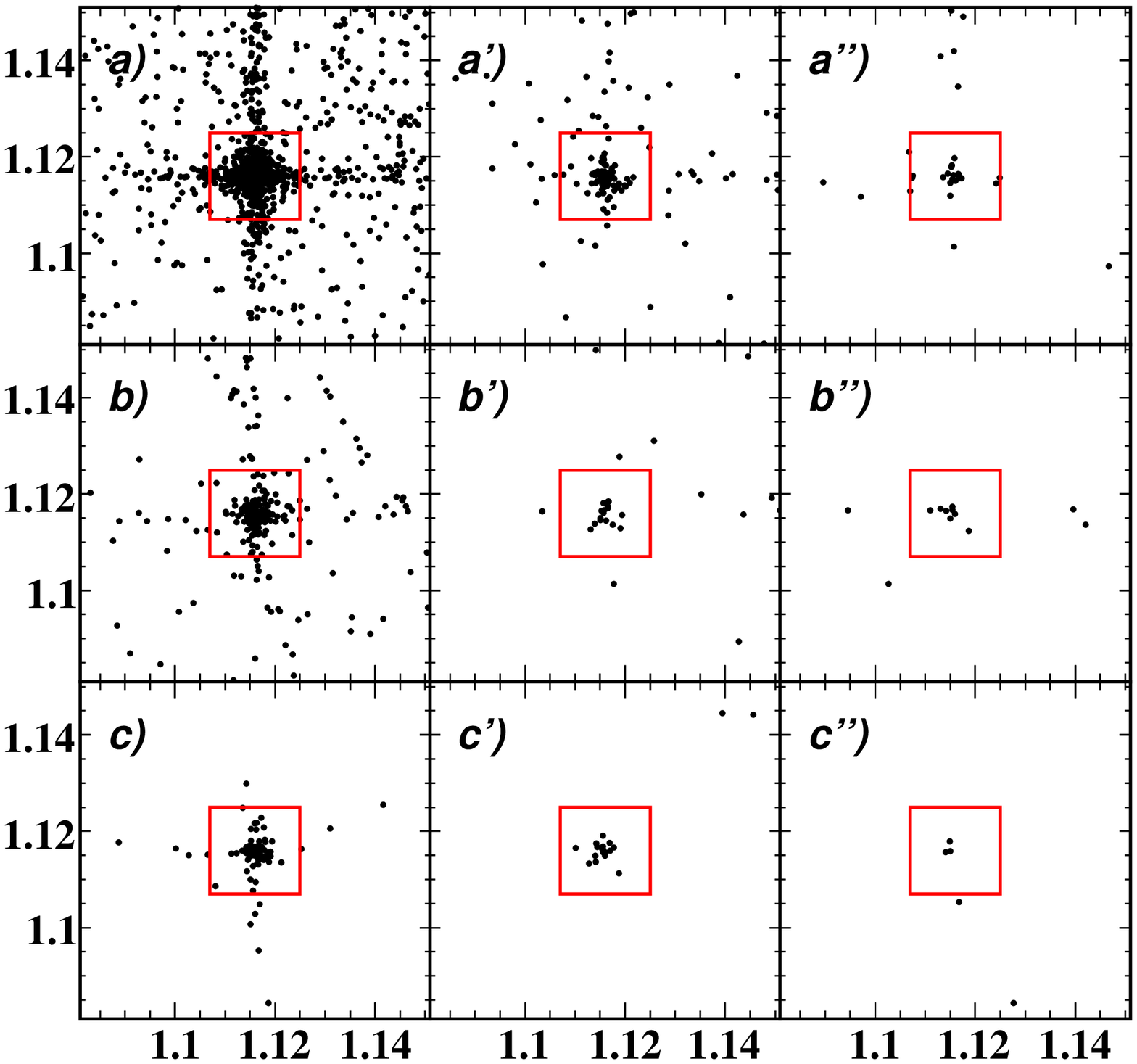}
  \put(-155,-2){\normalsize \bf \boldmath M(p$\pi^-$) (GeV/$c^2$)}
  \put(-250,85){\rotatebox{90}{\normalsize \bf \boldmath M($\bar p\pi^+$) (GeV/$c^2$)}}
  \caption{ \label{2D_1}
  Invariant mass of $p\pi^-$ versus $\bar p\pi^+$ distributions
  for $\Lambda\bar\Lambda\pi^+\pi^-$ [(a), (a') and (a'')],
  $\Lambda\bar\Lambda\pi^0$ [(b), (b') and (b'')], $\Lambda\bar\Lambda
  \eta$ [(c), (c') and (c'')]. The rectangle regions indicate signal regions.
  The figures on the left (middle, right) side: data at $\sqrt s
  =3.773$ [4.009, continuum (3.543, 3.554, 3.561, 3.600 and 3.650)] GeV.}
\end{figure}

For every final state, the invariant mass distributions of the reconstructed
intermediate states have the signal range determined from
Monte Carlo studies:
$\Lambda$ ($1.107\leq M(p\pi^-)\leq1.124$ GeV/$c^2$), 
$\pi^{0}$ ($115\leq M(\gamma\gamma)\leq150$ MeV/$c^2$),
$\eta$ ($515\leq M(\gamma\gamma)\leq569$ MeV/$c^2$),
$\Sigma^+$ ($1.164\leq M(p\gamma\gamma)\leq1.206$ GeV/$c^2$),
$\Sigma^0$ ($1.178\leq M(p\pi^-\gamma)\leq1.205$ GeV/$c^2$),
$\Xi^-$ ($1.305\leq M(p\pi^-\pi^-)\leq1.337$ GeV/$c^2$),
$\Xi^0$ ($1.281\leq M(p\pi^-\gamma\gamma)\leq1.330$ GeV/$c^2$).
For $\Xi^0\bar{\Xi^0}$, the selection of $\pi^{0}$ has
a looser requirement of $110\leq M(\gamma\gamma)\leq150$ MeV/$c^2$
due to the clean signal.
With regard to any of the unstable particles, the signal
range is about $3\sigma$ around the known mass of the particle,
and the sideband range (not shown) is approximately from $5\sigma$ to
$8\sigma$ at each side of the particle, where $\sigma$ is
the resolution determined by Monte Carlo simulation. In Fig. \ref{Inv_Mass},
the invariant mass distributions are shown for
each reconstructed intermediate state, for which the
events pass all the above selections.

\begin{figure}[htbp]
\includegraphics[width=8.6cm,height=11.6cm] {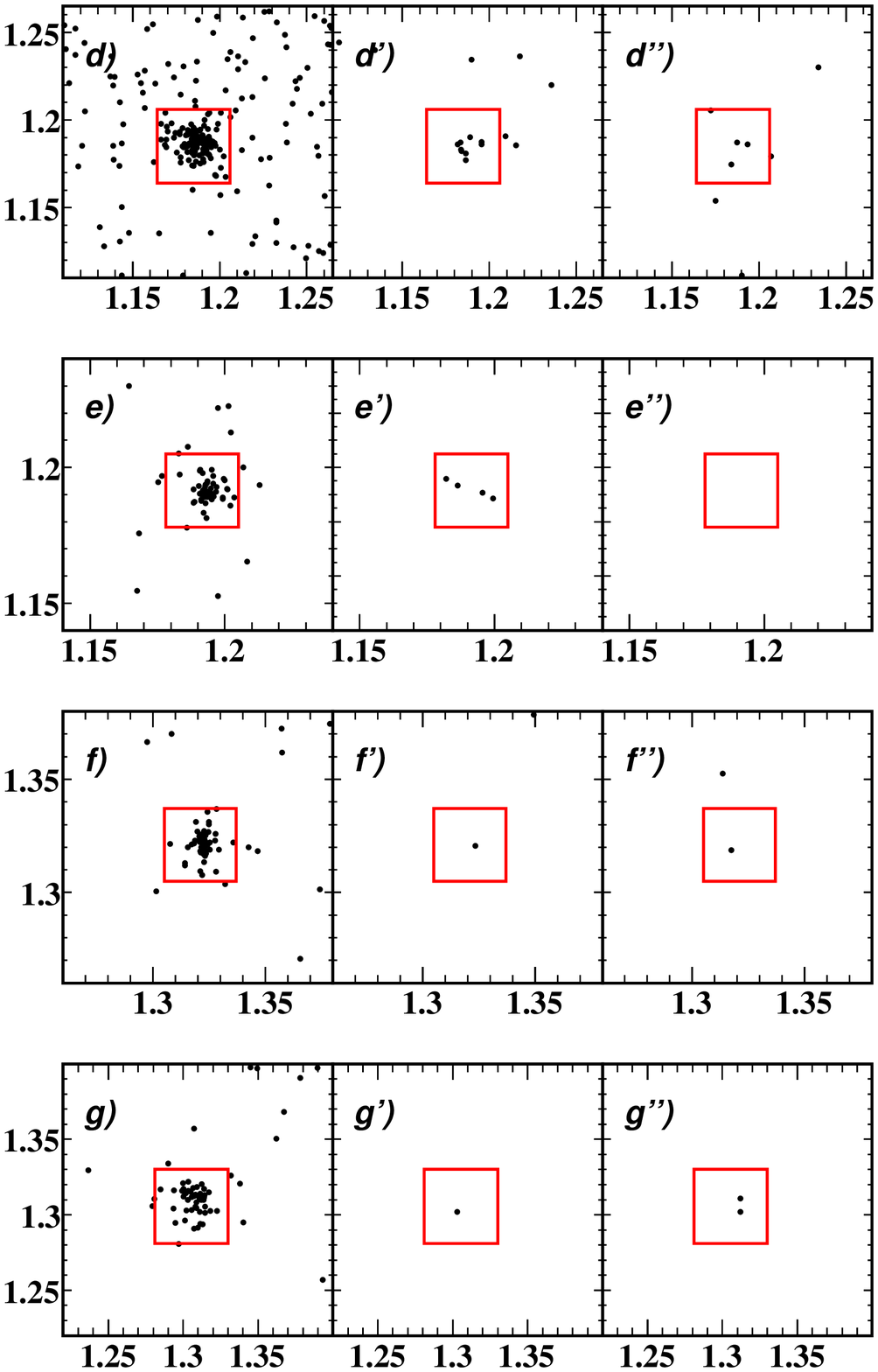}
  \put(-155,248){\normalsize \bf \boldmath M(p$\gamma\gamma$) (GeV/$c^2$)}
  \put(-155,166){\normalsize \bf \boldmath M(p$\pi^-\gamma$) (GeV/$c^2$)}
  \put(-156,84 ){\normalsize \bf \boldmath M(p$\pi^-\pi^-$) (GeV/$c^2$)}
  \put(-155,-2 ){\normalsize \bf \boldmath M(p$\pi^-\gamma\gamma$) (GeV/$c^2$)}
  \put(-250,75){\rotatebox{90}{\normalsize \bf \boldmath M($\bar p\pi^+\gamma\gamma$/$\bar p\pi^+\pi^+$/$\bar p\pi^+\gamma$/$\bar p\gamma\gamma$) (GeV/$c^2$)}}
  \caption{ \label{2D_2} Invariant mass of $p\gamma\gamma$,
  $p\pi^-\gamma$, $p\pi^-\pi^-$ or $p\pi^-\gamma\gamma$ versus $\bar
  p\gamma\gamma$, $\bar p\pi^+ \gamma$, $\bar p\pi^+\pi^+$ or $\bar
  p\pi^+\gamma\gamma$ distributions for $\Sigma^+\bar{\Sigma^-}$ [(d),
  (d') and (d'')], $\Sigma^0\bar{\Sigma^0}$ [(e), (e') and (e'')],
  $\Xi^-\bar{\Xi^+}$ [(f), (f') and (f'')], $\Xi^0\bar{\Xi^0}$ [(g),
  (g') and (g'')]. The rectangle regions indicate signal regions. The
  figures on the left (middle, right) side: data at $\sqrt s=3.773$
  [4.009, continuum (3.543, 3.554, 3.561, 3.600 and 3.650)] GeV. }
\end{figure}

For each mode studied, the signal selection region in the two
dimensional scatter plot is determined by Monte Carlo simulation. In
Figs.~\ref{2D_1} and \ref{2D_2}, the two dimensional scatter plots are
shown for each mode. The rectangle regions are the signal regions. For
final states of $\Lambda\bar\Lambda\pi^0$ and
$\Lambda\bar\Lambda\eta$, the plots show the distributions by
requiring $\pi^0$ and $\eta$ to be in the signal range. To determine
signal yields, the sideband events from $\pi^0$ and $\eta$ must be
removed. We first extract the number of events in one particle signal
range with the requirement that the other particle falls in its signal
and sideband range, defined to be 'signal' and 'sideband',
respectively, and then obtain the observed events $N_{obs}$ after
removing the normalized 'sideband' events from 'signal'.

\section{\boldmath Background Estimation}
Our foremost observable is the background-subtracted number
of the baryonic events inferred to be directly from $\psi(3770)$
and $\psi(4040)$ decays, $N^{S}_{\psi(3770)\to f}$ and
$N^{S}_{\psi(4040)\to f}$. For the data taken at the center-of-mass
energy of 3.773 GeV, the background contributions to the
baryonic final states come from continuum production
$e^+e^-\to q\bar q\to f$, $N_{q\bar q}^f(3.773)$, the initial
state radiative returns to $\psi(3686)$ and $J/\psi$ production
$e^+e^-\to\gamma^{ISR}\psi(3686)$ $(J/\psi)\to f$, $N_{\gamma\psi(3686)}^f(3.773)$
($N_{\gamma J/\psi}^f(3.773)$), and the misidentified $\psi(3770)$
direct decays mainly containing $D\bar D$ production,
$N_{D\bar D}^f(3.773)$. For the data taken at 
the center-of-mass energy of 4.009 GeV, the background contributions
to the baryonic final states come from continuum production
$e^+e^-\to q\bar q\to f$, $N_{q\bar q}^f(4.009)$, the initial
state radiative returns to $\psi(3770)$, $\psi(3686)$ and $J/\psi$
production $e^+e^-\to\gamma^{ISR}\psi(3770)$ $(\psi(3686)$, $J/\psi)\to f$,
$N_{ISR}^f(4.009)$, and the misidentified $\psi(4040)$
direct decays containing open charm (DD), hadronic (hadrons) and
radiative (gammaXYZ) production, $N_{DHG}^f(4.009)$.

Based on the Monte Carlo samples of $e^+e^-\to\gamma^{ISR}J/\psi$,
$\gamma^{ISR}\psi(3686)$, and $e^+e^-\to\psi(3770)\to D\bar D$
generated at the center-of-mass energy 3.773 GeV, the backgrounds of
$N_{\gamma\psi(3686)}^f(3.773)$, $N_{\gamma J/\psi}^f(3.773)$ and
$N_{D\bar D}^f(3.773)$ were studied by employing the similar analysis
strategy as described previously. Thus this part of background,
$N_{B}^f(3.773)$, is given by
\begin{equation}
\begin{split}
N_{B}^f(3.773) &= f_{\gamma\psi(3686)}\times N_{\gamma\psi(3686)}^f(3.773) \\
               &+ f_{\gamma J/\psi}\times N_{\gamma J/\psi}^f(3.773) \\
               &+ f_{D\bar D}\times N_{D\bar D}^f(3.773),
\label{N_bkg_psipp}
\end{split}
\end{equation}
where $f_{\gamma\psi(3686)}=f_{\gamma J/\psi}=1/1.5$ and $f_{D\bar D}=1/5.0$
are the scale factors for the Monte Carlo samples. The results of
$N_{B}^f(3.773)$ are listed in Table \ref{tab:sig_psipp}.  

Using the Monte Carlo samples of $e^+e^-\to\gamma^{ISR}J/\psi$,
$\gamma^{ISR}\psi(3686)$, $\gamma^{ISR}\psi(3770)$, and $e^+e^-\to\psi(4040)\to$
DD, hadrons, gammaXYZ generated at the center-of-mass energy 4.009 GeV,
the backgrounds of $N_{ISR}^f(4.009)$ and $N_{DHG}^f(4.009)$ are studied
by employing the similar analysis strategy as described previously.
Thus this part of background, $N_{B}^f(4.009)$, is given by
\begin{equation}
\begin{split}
N_{B}^f(4.009)& = f_{ISR}\times N_{ISR}^f(4.009) \\
              & + f_{DHG}\times N_{DHG}^f(4.009)
\label{N_bkg_psi4010}
\end{split}
\end{equation}
\noindent where $f_{ISR}=f_{DHG}=1/2.1$ are the scale factors for the Monte
Carlo samples. The results of $N_{B}^f(4.009)$ are listed
in Table~\ref{tab:sig_psi4040}.  

To estimate the largest background at center-of-mass energies of 3.773
and 4.009 GeV, the continuum production $e^+e^-\to\gamma^*\to q\bar
q\to f$, $N_{q\bar q}^f(3.773)$ and $N_{q\bar q}^f(4.009)$, the data
taken at center-of-mass energies of 3.542, 3.554, 3.561, 3.600 and
3.650 GeV, which have small contaminations from the $\psi(3686)$ lower end
intrinsic tail decays as well as of radiative returns to $J/\psi$
decays, is used. Hence $N_{q\bar q}^f(3.773/4.009)$ is obtained by
\begin{equation}
\begin{split}
 N_{q\bar q}^f(3.773/4.009) &= f_{co}^{3.773/4.009}\times N_{q\bar q}^f(3.650) \\ 
                      &= f_{co}^{3.773/4.009}\times[N_{obs}^f(3.650) \\
                      &-N_{B}^f(3.650)] \\
                      &= f_{co}^{3.773/4.009}\times[N_{obs}^f(3.650)\\
                      &-N_{\psi(3686)}^f(3.650)- N_{\gamma J/\psi}^f(3.650)],
\label{tab:N_qq}
\end{split}
\end{equation}
\noindent where $N_{obs}^f(3.650)$ is the observed number of baryonic final
state events in the continuum data taken at center-of-mass energies of
3.542, 3.554, 3.561, 3.600 and 3.650 GeV, from which we scale the
events from the first four energy points to the energy point of 3.650
GeV by considering the different efficiency and the assumed $1/s$
dependence of the cross section, $N_{\psi(3686)}^f(3.650)$ and
$N_{\gamma J/\psi}^f(3.650)$ are the number of baryonic final state
events from $\psi(3686)$ and $J/\psi$ decays, respectively. $N_{\gamma
J/\psi}^f(3.650)$ is obtained in the same way as previously described
but at the center-of-mass energy of 3.650
GeV. $N_{\psi(3686)}^f(3.650)$ is given by
$N_{\psi(3686)}^f(3.650)=\sigma_{\psi(3686)}^{3.650}\times
\mathcal{L}\times\epsilon_{\psi(3686)\to f}^{3.650}$, where
$\sigma_{\psi(3686)}^{3.650}$ is the cross section for $\psi(3686)$
production at the center-of-mass energy of 3.650 GeV
\cite{prd76_122002}, and $\epsilon_{\psi(3686)\to f}^{3.650}$ is the
baryonic final state event selection efficiency of events from
$\psi(3686)\to f$ at the center-of-mass energy of 3.650 GeV,
determined by Monte Carlo simulation.  The scaling factor,
$f_{co}^{3.773/4.009}$, is mode dependent and determined by the
integrated luminosities of the two data sets corrected for an assumed
$1/s$ dependence of the cross section, and accounts for the small
difference in efficiency between the $\psi(3770)$/$\psi(4040)$ data
and continuum data. The uncertainty of $f_{co}^{3.773/4.009}$, about
2.0\%-3.0\%, arises from the uncertainties in relative luminosity and
detection efficiencies at the two energy points. The results of
$N_{obs}^f(3.650)$, $N_{B}^f(3.650)$ and $f_{co}^{3.773/4.009}$ are
also listed in Table~\ref{tab:sig_psipp} and
Table~\ref{tab:sig_psi4040}.

\begin{table*}[htbp]
\begin{center}
\caption{ \label{tab:sig_psipp} For each mode $f$ the following
quantities are given: the number of observed events,
$N_{obs}^f(3.773)$, and background events, $N_{B}^f(3.773)$,
containing $N_{\gamma\psi(3686)}^f(3.773)$, $N_{\gamma
J/\psi}^f(3.773)$ and $N_{D\bar D}^f(3.773)$ in $\psi(3770)$ data; the
number of observed events, $N_{obs}^f(3.650)$, and background events,
$N_{B}^f(3.650)$, containing $N_{\psi(3686)}^f(3.650)$ and $N_{\gamma
J/\psi}^f(3.650)$ in continuum data; the scale factor
$f_{co}^{3.773}$; the number of events attributable to $\psi(3770)$
decay, $N^{S}_{\psi(3770)\to f}$, computed according to
Eq. (\ref{N_psipp}); the upper limits on the number of events for
$\psi(3770)$ baryonic decays including the systematic error (90\%
C.L.), $N^{up}_{\psi(3770)\to f}$; the detection efficiency
$\epsilon$; the relative systematic error coming from the uncertainty
in luminosity, intermediate state branching fractions, Monte Carlo
statistics and the total number of $\psi(3770)$ decays,
$\Delta_{sys}$; the branching fraction ${\mathcal B}_{\psi(3770)\to
f}$; and the branching fraction upper limits for $\psi(3770)$ decays
including the systematic errors (90\% C.L.), ${\mathcal B}^{up}$.}
\renewcommand{\arraystretch}{1.3} \scriptsize
\begin{tabular}{|c|c|c|c|c|c|c|c|c|c|c|c|} \hline
Mode & $N_{obs}^f(3.773)$ & $N_{B}^f(3.773)$ &
$N_{obs}^f(3.650)$ & $N_{B}^f(3.650)$ & $f_{co}^{3.773}$ &
$N^{S}_{\psi(3770)\to f}$ & $N^{up}_{\psi(3770)\to f}$ &
$\epsilon$ & $\Delta_{sys}$ &
${\mathcal B}_{\psi(3770)\to f}$ & ${\mathcal B}^{up}$ \\
f& & & & & & & & & & [$\times 10^{-4}$] & [$\times 10^{-4}$] \\ \hline 
$\Lambda\bar{\Lambda}\pi^+\pi^-$  &$844.0\pm33.6$&$5.2$&$14.2^{+5.6}_{-4.2}$&$0.1$&$45.27$&$ 200.6^{+193.1}_{-255.7}\pm42.0$&$481.2$&$0.1321$&$8.0$&$ 1.80^{+1.74}_{-2.30}\pm0.40$&$<4.7$ \\
$\Lambda\bar{\Lambda}\pi^0$       &$124.9\pm14.4$&$3.4$&$7.1^{+5.0}_{-2.2} $&$0.0$&$42.50$&$-180.3^{+94.6 }_{-213.0}\pm16.2$&$83.6 $&$0.1694$&$8.0$&$-1.28^{+0.67}_{-1.51}\pm0.15$&$<0.7$ \\
$\Lambda\bar{\Lambda}\eta$        &$74.0 \pm9.5 $&$0.9$&$3.0^{+3.6}_{-1.6} $&$0.0$&$44.76$&$-61.2 ^{+72.2 }_{-161.4}\pm7.9 $&$87.7 $&$0.1518$&$8.1$&$-1.22^{+1.44}_{-3.21}\pm0.19$&$<1.9$ \\
$\Sigma^+\bar{\Sigma^-}$          &$100.5\pm11.9$&$0.7$&$3.3^{+4.3}_{-1.7} $&$0.1$&$38.27$&$-22.7 ^{+66.1 }_{-165.0}\pm5.1 $&$96.0 $&$0.1975$&$8.0$&$-0.21^{+0.63}_{-1.56}\pm0.05$&$<1.0$ \\
$\Sigma^0\bar{\Sigma^0}$          &$43.5 \pm6.7 $&$0.0$&$0.0^{+2.2}_{-0.0} $&$0.0$&$38.69$&$ 43.5 ^{+6.7  }_{-85.4 }\pm5.8 $&$56.6 $&$0.1752$&$8.0$&$ 0.30^{+0.05}_{-0.58}\pm0.05$&$<0.4$ \\
$\Xi^-\bar{\Xi^+}$                &$48.5 \pm7.0 $&$0.0$&$0.5^{+2.8}_{-1.4} $&$0.0$&$41.74$&$ 27.6 ^{+58.9 }_{-117.1}\pm3.7 $&$119.7$&$0.1060$&$8.1$&$ 0.31^{+0.66}_{-1.32}\pm0.05$&$<1.5$ \\
$\Xi^0\bar{\Xi^0}$                &$43.5 \pm6.6 $&$1.3$&$2.0^{+3.2}_{-1.2} $&$0.0$&$40.13$&$-38.1 ^{+48.6 }_{-128.6}\pm5.6 $&$60.7 $&$0.0581$&$8.2$&$-0.80^{+1.03}_{-2.72}\pm0.14$&$<1.4$ \\
\hline
\end{tabular}
\end{center}
\end{table*}

\begin{table*}[htbp]
\begin{center}
\caption{ \label{tab:sig_psi4040} For each mode $f$ the following
quantities are given: the number of observed events,
$N_{obs}^f(4.009)$, and background events,
$N_{B}^f(4.009)$, containing $N_{ISR}^f(4.009)$,
$N_{DHG}^f(4.009)$ in $\psi(4040)$ data; the number of observed
events, $N_{obs}^f(3.650)$, and background events,
$N_{B}^f(3.650)$, containing $N_{\psi(3686)}^f(3.650)$
and $N_{\gamma J/\psi}^f(3.650)$ in continuum data; the scale factor
$f_{co}^{4.009}$; the number of events attributable to $\psi(4040)$
decay, $N^{S}_{\psi(4040)\to f}$, computed according to
Eq. (\ref{N_psipp}); the upper limits on the number of events for
$\psi(4040)$ baryonic decays including the systematic error (90\%
C.L.), $N^{up}_{\psi(4040)\to f}$; the detection efficiency
$\epsilon$; the relative systematic error coming from the
uncertainty in luminosity, intermediate state branching fractions,
Monte Carlo statistics and the number of $\psi(4040)$ decays,
$\Delta_{sys}$; the branching fraction ${\mathcal B}_{\psi(4040)\to
f}$; and the branching fraction upper limits for $\psi(4040)$ decays
including the systematic errors (90\% C.L.), ${\mathcal B}^{up}$.}
\renewcommand{\arraystretch}{1.3} \scriptsize
\begin{tabular}{|c|c|c|c|c|c|c|c|c|c|c|c|} \hline
Mode & $N_{obs}^f(4.009)$ & $N_{B}^f(4.009)$ &
$N_{obs}^f(3.650)$ & $N_{B}^f(3.650)$ & $f_{co}^{4.009}$ &
$N^{S}_{\psi(4040)\to f}$ & $N^{up}_{\psi(4040)\to f}$ &
$\epsilon$ & $\Delta_{sys}$ &
${\mathcal B}_{\psi(4040)\to f}$ & ${\mathcal B}^{up}$ \\
f& & & & & & & & & & [$\times 10^{-4}$] & [$\times 10^{-4}$] \\ \hline 
$\Lambda\bar{\Lambda}\pi^+\pi^-$  &$79.2\pm10.0       $&$20.0$&$14.2^{+5.6}_{-4.2}$&$0.1$&$7.69$&$-49.2^{+33.8}_{-44.2}\pm9.8$&$35.6$&$0.1492$&$9.9 $&$-3.57^{+2.45}_{-3.21}\pm0.79$&$<2.9$ \\
$\Lambda\bar{\Lambda}\pi^0$       &$14.5^{+4.1}_{-4.3}$&$0.5 $&$7.1^{+5.0}_{-2.2} $&$0.0$&$6.80$&$-34.3^{+15.5}_{-34.3}\pm3.0$&$12.6$&$0.1753$&$9.9 $&$-2.14^{+0.97}_{-2.14}\pm0.28$&$<0.9$ \\
$\Lambda\bar{\Lambda}\eta$        &$16.0^{+4.2}_{-4.3}$&$3.6 $&$3.0^{+3.6}_{-1.6} $&$0.0$&$7.38$&$-9.8 ^{+12.5}_{-26.9}\pm3.3$&$16.2$&$0.1674$&$9.9 $&$-1.60^{+2.06}_{-4.43}\pm0.57$&$<3.0$ \\
$\Sigma^+\bar{\Sigma^-}$          &$8.5^{+3.0}_{-3.2} $&$0.2 $&$3.3^{+4.3}_{-1.7} $&$0.1$&$4.92$&$-7.5 ^{+8.9 }_{-21.4}\pm1.5$&$11.0$&$0.1704$&$9.9 $&$-0.74^{+0.89}_{-2.14}\pm0.17$&$<1.3$ \\
$\Sigma^0\bar{\Sigma^0}$          &$4.0^{+3.2}_{-1.9} $&$0.0 $&$0.0^{+2.2}_{-0.0} $&$0.0$&$5.03$&$ 4.0 ^{+3.2 }_{-11.2}\pm0.5$&$8.9 $&$0.1537$&$9.9 $&$ 0.28^{+0.23}_{-0.79}\pm0.04$&$<0.7$ \\
$\Xi^-\bar{\Xi^+}$                &$1.0^{+2.2}_{-0.8} $&$0.0 $&$0.5^{+2.8}_{-1.4} $&$0.0$&$5.61$&$-1.8 ^{+8.2 }_{-15.7}\pm0.3$&$12.5$&$0.0941$&$9.9 $&$-0.21^{+0.94}_{-1.81}\pm0.04$&$<1.6$ \\
$\Xi^0\bar{\Xi^0}$                &$1.0^{+2.2}_{-0.8} $&$0.0 $&$2.0^{+3.2}_{-1.2} $&$0.0$&$5.36$&$-9.7 ^{+6.8 }_{-17.2}\pm1.3$&$7.0 $&$0.0490$&$10.0$&$-2.22^{+1.55}_{-3.93}\pm0.37$&$<1.8$ \\
\hline
\end{tabular}
\end{center}
\end{table*}

\section{\boldmath Results}
We assume that there is no interference between continuum production
and the $\psi(3770)$/$\psi(4040)$ resonance decay to the same baryonic
final state. To obtain the number of baryonic final state events
from $\psi(3770)$/$\psi(4040)$ direct decays, $N^{S}_{\psi(3770)/\psi(4040)\to f}$,
we define $N^{S}_{\psi(3770)/\psi(4040)\to f}$ as 
\begin{equation}
\begin{split}
N^{S}_{\psi(3770)/\psi(4040)\to f} &= N_{obs}^f(3.773/4.009) \\
                                   &-N_{B}^f(3.773/4.009) \\
                                   &-N_{q\bar q}^f(3.773/4.009),  
\label{N_psipp}
\end{split}
\end{equation}
\noindent where $N_{obs}^f(3.773/4.009)$ is the observed number of
baryonic final state events in the $\psi(3770)$/$\psi(4040)$ data
taken at the center-of-mass energy of 3.773/4.009 GeV. Since no
statistically significant extra signal is observed, 90\% C.L. upper
limits on the number of events for $\psi(3770)$/$\psi(4040)$ baryonic
decays are computed with systematic errors included for each mode,
$N^{up}_{\psi(3770)/\psi(4040)\to f}$, by assuming that they follow a
Gaussian distribution and considering only physical (positive)
values. The relative systematic errors related to
$N^{S}_{\psi(3770)/\psi(4040)\to f}$ include the independent and
common systematic errors.  The independent systematic errors depend on
energy points and consist of uncertainty in background subtraction
(0.0\%-20.2\%) and uncertainty in invariant mass spectrum fit
(0.0\%-5.1\%).  The common systematic errors do not depend on the
energy point and include the uncertainty in invariant mass requirement
for unstable particles (0.8\%-3.4\%) and also the detector performance
related quantities: charged particle tracking (1.0\% per track),
photon selection (1.0\% per photon), $\pi$/p/$\bar p$ identification
(1.0\%/1.0\%/2.0\%), vertex and secondary vertex fit (1.0\% each) and
kinematic fit (1.0\%-2.0\%). Some of the modes have tiny component
resonant submodes, however, the efficiencies do not differ by
much. Monte Carlo samples are generated with a phase space model, and
the difference of the efficiencies with and without intermediate
states is taken as the systematic error (0.3\%). Another systematic
error for baryon pairs comes from the angular distribution, which is
described by $1+\alpha cos^2\theta$, with $\theta$ being the polar
angle. Monte Carlo samples are generated for $\alpha = 0$ and $\alpha
= 1$, and the difference of efficiency between $\alpha = 0$ and
$\alpha = 1$ is taken as the systematic error (9.2\%-10.9\%).  The
results of $N_{obs}^f(3.773/4.009)$, $N^{S}_{\psi(3770)/\psi(4040)\to
f}$ and $N^{up}_{\psi(3770)/\psi(4040)\to f}$ are also shown in Tables
~\ref{tab:sig_psipp} and \ref{tab:sig_psi4040}.

The number of baryonic final state events from
$\psi(3770)$/$\psi(4040)$ decays, $N^{S}_{\psi(3770)/\psi(4040)\to f}$
is corrected by the detection efficiency, $\epsilon$ and the related branching
fractions for the intermediate state decays, $B_f$, and normalized to
the total number of the $\psi(3770)$/$\psi(4040)$ decays to obtain the
branching fraction.
\begin{equation}
\begin{split}
\mathcal{B}_{\psi(3770)/\psi(4040)\to f}  = 
\frac{N^{S}_{\psi(3770)/\psi(4040)\to f}}
{\epsilon\times B_f\times N_{\psi(3770)/\psi(4040)}},
\end{split}
\end{equation} 
The upper limit on the branching fraction can be
obtained by 
\begin{equation}
\begin{split}
\mathcal{B}^{up}  = 
\frac{N^{up}_{\psi(3770)/\psi(4040)\to f}}
{\epsilon\times B_f\times N_{\psi(3770)/\psi(4040)}\times(1-\Delta_{sys})},
\end{split}
\end{equation} 
\noindent where the detection efficiency for $\psi(3770)$/$\psi(4040)$ baryonic
decays, $\epsilon$ is estimated by using the Monte Carlo
simulation of the BESIII detector based on the KKMC \cite{KKMC}
and BesEvtGen \cite{EvtGen} generators. The detection efficiencies
in Tables \ref{tab:sig_psipp} and \ref{tab:sig_psi4040} do not include
the branching fractions for intermediate state decays. $B_f$ is the
branching fraction \cite{pdg2012} for the intermediate state decays for
each mode. $N_{\psi(3770)/\psi(4040)}$ is the number of
$\psi(3770)$/$\psi(4040)$ decays and is determined by
$N_{\psi(3770)/\psi(4040)}=\sigma_{\psi(3770)/\psi(4040)}^{Born-level}
\times\mathcal{L}_{\psi(3770)/\psi(4040)}\times(1+\delta)_{ISR}$, where
$\sigma_{\psi(3770)/\psi(4040)}^{Born-level}=(9.93\pm0.77)/(6.2\pm0.6)$ nb
is the Born-level cross section of $\psi(3770)$/$\psi(4040)$
at $\sqrt s=3.773$ and 4.009 GeV obtained by the relativistic
Breit-Wigner formula
with the $\psi(3770)$ and $\psi(4040)$ resonance parameters \cite{pdg2012};
$\mathcal{L}_{\psi(3770)/\psi(4040)}$ is the integrated luminosity
for $\psi(3770)$/$\psi(4040)$ data and $(1+\delta)_{ISR}=0.718/0.757$
is the radiative correction factor, obtained from the KKMC \cite{KKMC}
generator with the $\psi(3770)$ and $\psi(4040)$ resonance parameters
\cite{pdg2012} input. $\Delta_{sys}$ is the relative systematic error only
involving the uncertainty in the integrated luminosity (1.1\%),
the intermediate state branching fractions (0.8\%-1.2\%),
the Monte Carlo statistics (0.9\%-2.0\%) and a common uncertainty
of 7.8\%/9.7\% due to the number of $\psi(3770)$/$\psi(4040)$
decays arising from the uncertainty in $\psi(3770)$/$\psi(4040)$
resonance parameters. We give the branching fractions
$\mathcal{B}_{\psi(3770)/\psi(4040)\to f}$ and the upper limits on
branching fractions $\mathcal{B}^{up}$ of $\psi(3770)$/$\psi(4040)$
baryonic decays for each mode in Tables~ \ref{tab:sig_psipp} and
\ref{tab:sig_psi4040}. Since the available continuum data is
limited, the dominant error on each of the seven branching fractions
is from the continuum subtraction.

\section{\boldmath Summary}
In summary, using 2.9 fb$^{-1}$ of data taken at $\sqrt{s}=3.773$ GeV,
482 pb$^{-1}$ of data taken at $\sqrt{s}=4.009$ GeV, 23 pb$^{-1}$
of data taken at $\sqrt{s}=3.542$, 3.554, 3.561 and 3.600 GeV and
44 pb$^{-1}$ of data taken at $\sqrt{s}=3.650$ GeV collected with the
BESIII detector at the BEPCII collider, searches for seven baryonic
decays of $\psi(3770)$ and $\psi(4040)$ are presented; most are the
first searches. The upper limits on the branching fractions for
$\psi(3770)$ and $\psi(4040)$ baryonic decays are set at the 90\%
C.L.. The sum of the seven branching fractions are
$(-1.11^{+2.72}_{-5.46}\pm0.72)\times10^{-4}$ and
$(-1.02^{+0.39}_{-0.76}\pm0.15)\times10^{-3}$, and the corresponding
upper limits at the 90\% C.L. are $4.0\times10^{-4}$ and
$3.1\times10^{-4}$ for the seven baryonic decays of $\psi(3770)$ and
$\psi(4040)$, respectively. Although this study, together with
previous studies on searching for exclusive non-$D\bar D$, provide
useful information for understanding the nature of $\psi(3770)$, the
large non-$D\bar D$ component still remains a puzzle. A fine energy
scan over $\psi(3770)$ and $\psi(4040)$ resonances would be very
helpful for obtaining the lineshape of exclusive non-$D\bar D$
processes, and help determine whether the processes exist or not.

\section{\boldmath Acknowledgement} 
The BESIII collaboration thanks the staff of BEPCII and the computing center for their strong support. This work is supported in part by the Ministry of Science and Technology of China under Contract No. 2009CB825200, 2009CB825204; National Natural Science Foundation of China (NSFC) under Contracts Nos. 10625524, 10821063, 10825524, 10835001, 10935007, 11125525, 11235011; Joint Funds of the National Natural Science Foundation of China under Contracts Nos. 11079008, 11179007; the Chinese Academy of Sciences (CAS) Large-Scale Scientific Facility Program; CAS under Contracts Nos. KJCX2-YW-N29, KJCX2-YW-N45; 100 Talents Program of CAS; German Research Foundation DFG under Contract No. Collaborative Research Center CRC-1044; Istituto Nazionale di Fisica Nucleare, Italy; Ministry of Development of Turkey under Contract No. DPT2006K-120470; U. S. Department of Energy under Contracts Nos. DE-FG02-04ER41291, DE-FG02-05ER41374, DE-FG02-94ER40823; U.S. National Science Foundation; University of Groningen (RuG) and the Helmholtzzentrum fuer Schwerionenforschung GmbH (GSI), Darmstadt; WCU Program of National Research Foundation of Korea under Contract No. R32-2008-000-10155-0

\end{document}